\documentclass[twocolumn,prl,aps,superscriptaddress,showpacs]{revtex4}
\input{graphicx}
\def\Journal#1#2#3#4{{#1} {\bf #2}, #3 (#4)}
\def\EPJA{{\rm Eur. Phys. J.} A}

\def\JPG{{\rm J. Phys.} G}
\def\MPLA{{\rm Mod. Phys. Lett.} A}

\def\NIMA{{\rm Nucl. Instrum. Methods} A}
\def\NPA{{\rm Nucl. Phys.} A}

\def\PLB{{\rm Phys. Lett.}  B}
\def\PR{\rm Phys. Rev.}
\def\PRL{\rm Phys. Rev. Lett.}
\def\PRD{{\rm Phys. Rev.} D}
\def\PRC{{\rm Phys. Rev.} C}

\begin{document}
\title{Angular dependence of recoil proton polarization in high-energy $\gamma d \to p n$}

\author{X.~Jiang}
\affiliation{Rutgers, The State University of New Jersey, Piscataway, New Jersey 08854}

\author{J.~Arrington}
\affiliation{Argonne National Laboratory, Argonne, Illinois 60439}

\author{F.~Benmokhtar}
\affiliation{Rutgers, The State University of New Jersey, Piscataway, New Jersey 08854}

\author{A.~Camsonne}
\affiliation{Universit\'e Blaise Pascal/IN2P3, F-63177 Aubi\`ere, France}

\author{J.P.~Chen}
\affiliation{Thomas Jefferson National Accelerator Facility, Newport News, Virginia
23606}

\author{S.~Choi}
\affiliation{Temple University, Philadelphia, Pennsylvania 19122}

\author{E.~Chudakov}
\affiliation{Thomas Jefferson National Accelerator Facility, Newport News, Virginia
23606}

\author{F.~Cusanno}
\affiliation{INFN, Sezione Sanit\'a and Istituto Superiore di Sanit\'a, Laboratorio di
Fisica, I-00161 Rome, Italy}

\author{A.~Deur}
\affiliation{University of Virginia, Charlottesville, Virginia 22901}

\author{D.~Dutta}
\affiliation{Massachusetts Institute of Technology, Cambridge, Massachusetts 02139}

\author{F.~Garibaldi}
\affiliation{INFN, Sezione Sanit\'a and Istituto Superiore di Sanit\'a, Laboratorio di
Fisica, I-00161 Rome, Italy}

\author{D.~Gaskell}
\affiliation{Thomas Jefferson National Accelerator Facility, Newport News, Virginia
23606}

\author{O.~Gayou}
\affiliation{Massachusetts Institute of Technology, Cambridge, Massachusetts 02139}

\author{R.~Gilman}
\affiliation{Rutgers, The State University of New Jersey, Piscataway, New Jersey 08854}
\affiliation{Thomas Jefferson National Accelerator Facility, Newport News, Virginia
23606}

\author{C.~Glashauser}
\affiliation{Rutgers, The State University of New Jersey, Piscataway, New Jersey 08854}

\author{D.~Hamilton}
\affiliation{University of Glasgow, Scotland}

\author{O.~Hansen}
\affiliation{Thomas Jefferson National Accelerator Facility, Newport News, Virginia
23606}

\author{D.W.~Higinbotham}
\affiliation{Thomas Jefferson National Accelerator Facility, Newport News, Virginia
23606}

\author{R.J.~Holt}
\affiliation{Argonne National Laboratory, Argonne, Illinois 60439}

\author{C.W.~de Jager}
\affiliation{Thomas Jefferson National Accelerator Facility, Newport News, Virginia
23606}

\author{M.K.~Jones}
\affiliation{Thomas Jefferson National Accelerator Facility, Newport News, Virginia
23606}

\author{L.J.~Kaufman}
\affiliation{University of Massachusetts, Amherst, Massachusetts 01003}

\author{E.R.~Kinney}
\affiliation{University of Colorado, 390 UCB, Boulder, Colorado 80309}

\author{K.~Kramer}
\affiliation{College of William and Mary, Williamsburg, Virginia 23187}

\author{L.~Lagamba}
\affiliation{INFN, Sezione Sanit\'a and Istituto Superiore di Sanit\'a, Laboratorio di
Fisica, I-00161 Rome, Italy}

\author{R.~de Leo}
\affiliation{INFN, INFN/Sezione Baril, Bari, Italy}

\author{J.~Lerose}
\affiliation{Thomas Jefferson National Accelerator Facility, Newport News, Virginia
23606}

\author{D.~Lhuillier}
\affiliation{DAPNIA, Saclay, France}

\author{R.~Lindgren}
\affiliation{University of Virginia, Charlottesville, Virginia 22901}

\author{N.~Liyanage}
\affiliation{University of Virginia, Charlottesville, Virginia 22901}

\author{K.~McCormick}
\affiliation{Rutgers, The State University of New Jersey, Piscataway, New Jersey 08854}

\author{Z.-E.~Meziani}
\affiliation{Temple University, Philadelphia, Pennsylvania 19122}

\author{R.~Michaels}
\affiliation{Thomas Jefferson National Accelerator Facility, Newport News, Virginia
23606}

\author{B.~Moffit}
\affiliation{College of William and Mary, Williamsburg, Virginia 23187}

\author{P.~Monaghan}
\affiliation{Massachusetts Institute of Technology, Cambridge, Massachusetts 02139}

\author{S.~Nanda}
\affiliation{Thomas Jefferson National Accelerator Facility, Newport News, Virginia
23606}

\author{K.D.~Paschke}
\affiliation{University of Massachusetts, Amherst, Massachusetts 01003}

\author{C.F.~Perdrisat}
\affiliation{College of William and Mary, Williamsburg, Virginia 23187}

\author{V.~Punjabi}
\affiliation{Norfolk State University, Norfolk, Virginia 23504}

\author{I.A.~Qattan}
\affiliation{Argonne National Laboratory, Argonne, Illinois 60439}
\affiliation{Northwestern University, Evanston, Illinois 60208}

\author{R.D.~Ransome}
\affiliation{Rutgers, The State University of New Jersey, Piscataway, New Jersey 08854}

\author{P.E.~Reimer}
\affiliation{Argonne National Laboratory, Argonne, Illinois 60439}

\author{B.~Reitz}
\affiliation{Thomas Jefferson National Accelerator Facility, Newport News, Virginia
23606}

\author{A.~Saha}
\affiliation{Thomas Jefferson National Accelerator Facility, Newport News, Virginia
23606}

\author{E.C.~Schulte}
\affiliation{Argonne National Laboratory, Argonne, Illinois 60439}

\author{R.~Sheyor}
\affiliation{University of Tel Aviv, Tel Aviv, Israel}

\author{K.~Slifer}
\affiliation{Temple University, Philadelphia, Pennsylvania 19122}

\author{P.~Solvignon}
\affiliation{Temple University, Philadelphia, Pennsylvania 19122}

\author{V.~Sulkosky}
\affiliation{College of William and Mary, Williamsburg, Virginia 23187}

\author{G.M.~Urciuoli}
\affiliation{INFN, Sezione Sanit\'a and Istituto Superiore di Sanit\'a, Laboratorio di
Fisica, I-00161 Rome, Italy}

\author{E.~Voutier}
\affiliation{Institut des Sciences Nucleaires, Grenoble, France}

\author{K.~Wang}
\affiliation{University of Virginia, Charlottesville, Virginia 22901}

\author{K.~Wijesooriya}
\affiliation{Argonne National Laboratory, Argonne, Illinois 60439}

\author{B.~Wojtsekhowski}
\affiliation{Thomas Jefferson National Accelerator Facility, Newport News, Virginia
23606}

\author{L.~Zhu}
\affiliation{Massachusetts Institute of Technology, Cambridge, Massachusetts 02139}

\collaboration{The Jefferson Lab Hall A Collaboration}

\date{\today}

\begin{abstract}
We measured the angular dependence of the three recoil proton 
polarization components in two-body photodisintegration 
of the deuteron at a photon energy of 2 GeV.
These new data provide a benchmark for
calculations based on quantum chromodynamics.
Two of the five existing models have made predictions of polarization
observables.
Both explain the longitudinal polarization transfer satisfactorily.
Transverse polarizations are not well described, but
suggest isovector dominance.
\end{abstract}


\maketitle


An important question to nuclear physics is whether
one can understand {\em exclusive} nuclear reactions starting from
quantum chromodynamics (QCD), or at least QCD-inspired quark models.
Several studies of nuclei \cite{alexa99,abbott00,benmokhtar05,rvachev05}
at high momentum transfer, but in elastic or quasifree kinematics, 
have shown that the role of explicit quark degrees of freedom is 
subtle and elusive; 
even data that probe short ranges can be understood with 
hadronic theories with the underlying interactions
determined from the measured $NN$ force and other inputs.
Deuteron photodisintegration is unique among exclusive reactions, 
in that it has been measured at both large momentum
transfer and large energies.
Experiments \cite{ne8,ne17,89012,96003,99008,clas}
have shown that the cross section in the GeV region scales,
approximately following the constituent counting rules \cite{ccr,rossi05},
as would be expected if the underlying dynamics involved quark 
degrees of freedom.
The large energies involve
a sum over large
numbers of baryonic resonances, leading naturally to the 
idea of using quark degrees of freedom to explain the data,
although perturbative QCD is not expected to apply \cite{gg02}.

Several quark model calculations \cite{rna,qgs,hrm,ar,ljd}
give competing approximate explanations of the photodisintegration
cross sections.
All but one are based on the idea that the high energy photon is absorbed
by a pair of quarks being interchanged between the two nucleons;
the quark-gluon string (QGS) model \cite{qgs} uses Regge theory to 
evaluate 3-quark exchange.
The models calibrated by the measured $NN$ force \cite{qgs,hrm,ljd}
tend to better reproduce the cross sections than the models which
evaluate the quark-exchange diagram approximately \cite{rna,ar}. 
The reason \cite{ar} is that the data scale better with energy 
than one would expect for the measured energy range.

Polarization observables provide potentially stricter tests of 
the underlying dynamics \cite{qgspol,hrmpol}.
For example, the $\Sigma$ (linearly polarized photon) asymmetry,
measured up to 1.6 GeV, is important for constraining the isoscalar 
vs.\ isovector nature of the photon coupling at high energy \cite{qgspol}.
For the recoil proton polarizations, measured at $\theta_{cm}$ = 90$^\circ$ 
up to 2.4 GeV \cite{wijesooriya01},
the induced polarization $p_y$ vanishes, but the polarization
transfers $C_{x'}$ and $C_{z'}$ do not, although they are consistent
with a slow approach to zero.
Thus, while hadron helicity conservation (HHC) \cite{hhc} does not hold,
a slow approach to HHC cannot be ruled out,
but it is not expected \cite{gpr}.
Very small induced polarizations, consistent with the data, were predicted
by the hard rescattering (HR) model \cite{hrmpol}, based on modeled $pn$ 
helicity amplitudes.
Here, we present new data at large energy and four-momentum transfer,
to further test the theoretical models.
These data are the only polarization angular distribution in deuteron
photodisintegration measured significantly above 1 GeV.


The experiment (E00-007) ran in Hall A of the Thomas Jefferson National
Accelerator Facility (JLab) \cite{hallanim}.
A strained GaAs crystal produced the longitudinally polarized electron beam.
The beam helicity state was flipped pseudo-randomly at 30 Hz.
Beam charge asymmetries between the two helicity states were negligible.
The Hall A M{\o}ller polarimeter determined the average beam polarization,
$P_e$, 
to be 76\% $\pm$ 0.3\% (statistics) $\pm$ 3.0\% (systematics).

Circularly polarized bremsstrahlung photons were generated 
when the 2.057-GeV electron beam impinged on a 
copper radiator with a thickness of 6\% 
radiation length, located 
20 cm upstream of the center of a 15-cm liquid deuterium target.
The ratio of the photon polarization $P_{\gamma}$ to $P_e$
is calculable \cite{aalm};
here $P_{\gamma}/P_e \approx 99.5 \%$. 

Recoil protons from the target were detected in the Hall A
left high resolution spectrometer (HRSL) at the five 
kinematic settings listed in Table~\ref{tab:gdkin}.
The scattering angles, momentum, and interaction position
at the target were calculated from trajectories measured with
Vertical Drift Chambers (VDCs) located in the focal plane.
Two planes of plastic scintillators provided triggering and flight time
information for particle identification.
The incident photon energy was reconstructed from the scattered proton
momentum and angle using two-body photodisintegration kinematics.
Only events between the bremsstrahlung endpoint and the
pion production threshold were used in the analysis.

The final element in the detector stack was the focal plane 
polarimeter (FPP)~\cite{punjabi05}.
To improve its efficiency, we configured the polarimeter with
a dual analyzer system, 
as in \cite{rcs05}.
The VDC chambers, a 44-cm thick polyethylene (CH$_2$) analyzer, and 
front straw chambers constituted the first polarimeter, 
while the front straw chambers, a 49.5-cm carbon analyzer, 
and rear straw chambers constituted a second, independent polarimeter.

\begin{table}[h]
\begin{center}
\caption{\label{tab:gdkin} Kinematics of the data and FPP
parameters. A 2 GeV photon energy corresponds to a center
of mass energy of $W$ = 3.3 GeV.
}
\begin{tabular}{|cc||cc||ccc|}
\hline
$\langle E_\gamma \rangle$ & $\langle\theta^{~p}_{cm}\rangle$ & $\langle p_p\rangle$ & $\langle\theta^{~p}_{lab}\rangle$ & spin & FPP & elastic ${\vec e} p$ \\ 
   GeV                     &                            &  GeV/c &        & precession & analyzer & calibration \\ \hline
   2.01                   &  36.9$^\circ$  & 2.41 & 19.8$^\circ$ &  226.5$^\circ$ &  dual  & yes \\ \hline
   2.00                   &  52.9$^\circ$  & 2.24 & 29.1$^\circ$ &  209.6$^\circ$ &  dual  & yes \\ \hline
   1.99                   &  69.8$^\circ$  & 2.01 & 39.7$^\circ$ &  191.9$^\circ$ &  dual  & yes \\ \hline
   1.97                   &  89.8$^\circ$  & 1.68 & 53.9$^\circ$ &  167.1$^\circ$ &  dual  & yes \\ \hline
   1.94                   &  109.5$^\circ$ & 1.35 & 70.6$^\circ$ &  144.2$^\circ$ & carbon & no  \\ \hline
\end{tabular}
\end{center}
\end{table}

The HRSL spectrometer transverse angle and transverse position resolutions (FWHM) 
are 2.6 mrad and 4 mm, respectively~\cite{hallanim}. 
Events originating from the target windows
were eliminated by cuts on the interaction position. 
In addition to cuts based on particle's position and direction at the target, 
a set of 2-dimensional profile cuts 
on the particle position and angle correlations at the 
focal plane were applied to further eliminate background events.
The focal plane profiles were obtained from continuous-momentum spectra
taken without the bremsstrahlung radiator.
The subtraction of background processes from electro-disintegration 
used the same techniques as
for previous photodisintegration cross-section \cite{ne8,ne17,89012,96003,99008}
and polarization \cite{wijesooriya01,wijesooriya02} measurements.

The transverse proton polarization components at the focal plane
lead to azimuthal asymmetries in re-scattering in
the analyzer due to spin-orbit interactions. 
The alignment of the FPP chambers was determined with straight-through 
trajectories, with the analyzers removed.
Spin transport in the spectrometer was taken into account 
using a magnetic model calculation. 
The induced (transferred) polarization was determined by a 
maximum likelihood method using the sum (difference) of the 
azimuthal distributions corresponding to the two beam helicity states.
If necessary, the proton polarization components obtained in the laboratory 
frame are transformed into the center-of-mass frame.
Previous experiments
\cite{jones00,wijesooriya01,gayou01,gayou02,wijesooriya02,punjabi05,hallahe403,rcs05,ndelta05}
used the same procedures.

\begin{table}[htbp]
\begin{center}
\caption{\label{tab:fom} 
The figure of merit (FOM), $\epsilon A^2$, efficiency times
analyzing power squared, in \%,
and the proton form factor ratios obtained from the $ep$ calibration runs.
The FOM relative statistical uncertainties are $\approx$15\%.
Statistical uncertainties dominate the form factor ratio uncertainty.
$T_{p_{analy}}$, in GeV, is the proton kinetic energy at the
mid-point of each analyzer.
}
\begin{tabular}{|c|c||c|c||c|c||c|c|}
\hline
 $p_p^{init}$  & $T_p^{init}$ & \multicolumn{2}{|c||}{CH$_2$} & 
\multicolumn{2}{|c||}{Carbon} & $Q^2$ & $\mu G_E/G_M$  \\ \hline
 GeV/$c$  & GeV &  $T_{p_{analy}}$  & FOM & $T_{p_{analy}}$
& FOM  & GeV$^2$ &  \\ \hline
 2.41   & 1.65  & 1.60  &  0.41  & 1.45  & 0.52 & 3.10 & 0.65$\pm$0.11 \\ \hline
 2.23   & 1.49  & 1.45  &  0.48  & 1.29  & 0.69 & 2.78 & 0.60$\pm$0.12 \\ \hline
 2.00   & 1.28  & 1.24  &  0.70  & 1.08  & 0.92 & 2.38 & 0.79$\pm$0.11 \\ \hline
 1.68   & 0.99  & 0.94  &  0.99  & 0.78  & 1.53 & 1.85 & 0.72$\pm$0.05 \\ \hline
\end{tabular}
\end{center}
\end{table}

Calibration runs used  $\vec{e}p \rightarrow e\vec{p}$
elastic scattering at 4 GeV for four of the five
spectrometer momentum settings of this experiment.
Polarization transfer in elastic $\vec{e}p$ scattering
determines both the ratio of the proton electromagnetic 
form factors~\cite{akre,dombey,akre2,acg} and the 
FPP calibration, after accounting for beam polarization and
spin transport through the spectrometer; see Table~\ref{tab:fom}.
Our measurements agree well with previous data for 
analyzing powers of carbon and CH$_2$ analyzers~\cite{punjabi05,rcs05,mcn,pomme}
and for the form factor ratio~\cite{punjabi05}.
For the $\gamma d$ data at $\theta^p_{cm}=109.5^\circ$, only the carbon analyzer 
was used due to the low outgoing proton momentum; 
the analyzing power was taken from an
earlier $\vec{e}p$ calibration run with the same FPP set up~\cite{ndelta05}. 
For the four kinematics with dual analyzers, recoil polarizations 
were consistent between the two analyzers within uncertainties, 
and the weighted averages are given as the final results in Table~\ref{tab:cmave}.

\begin{table}[h]
\begin{center}
\caption{\label{tab:cmave} Center-of-mass frame proton recoil 
polarization components, with statistical and systematical uncertainties.
}
\vspace{0.2cm}
\begin{tabular}{|c||c|c|c|}
\hline
 $\theta^p_{cm}$ &     $p_y$          &     $C^\prime_x$    &  $C^\prime_z$     \\ \hline
 36.9$^\circ$  & -0.301 $\pm$ 0.053 &  -0.170 $\pm$ 0.041 & 0.654 $\pm$ 0.056 \\
               &        $\pm$ 0.029 &         $\pm$ 0.020 &       $\pm$ 0.051 \\ \hline
 52.9$^\circ$  & -0.209 $\pm$ 0.041 &  -0.205 $\pm$ 0.040 & 0.573 $\pm$ 0.071 \\ 
               &        $\pm$ 0.052 &         $\pm$ 0.031 &       $\pm$ 0.092 \\ \hline
 69.8$^\circ$  &  0.008 $\pm$ 0.033 &  -0.228 $\pm$ 0.045 & 0.835 $\pm$ 0.108 \\
               &        $\pm$ 0.039 &         $\pm$ 0.033 &       $\pm$ 0.116 \\ \hline
 89.8$^\circ$  & -0.090 $\pm$ 0.049 &   0.065 $\pm$ 0.074 & 0.453 $\pm$ 0.162 \\ 
               &        $\pm$ 0.045 &         $\pm$ 0.034 &       $\pm$ 0.065 \\ \hline
 109.5$^\circ$ &  0.226 $\pm$ 0.073 &   0.316 $\pm$ 0.082 & 0.001 $\pm$ 0.119 \\ 
               &        $\pm$ 0.053 &         $\pm$ 0.035 &       $\pm$ 0.037 \\ \hline
\end{tabular}
\end{center}
\end{table}
 
There are several systematic uncertainties.
The statistical uncertainties of the measured analyzing powers 
dominate over the beam polarization uncertainty.
Spectrometer offsets also contribute.
Potential geometrical 
biases are eliminated by requiring that all possible secondary 
scattering proton azimuthal angles ($\phi_{FPP}$) fall into the 
boundaries of the polarimeter.
The induced polarization $p_y$ in $ep$ elastic scattering vanishes --
neglecting small effects from two-photon exchange --
allowing a direct measurement of the false asymmetries in the polarimeter.
The $ep$ induced polarization measurements were all consistent with vanishing,
so the statistical accuracies of $p_y$ in the $ep$ calibration ($\le 0.04$),
were assigned as the false asymmetry systematic uncertainties of the induced 
polarization in deuteron photodisintegration.
For the polarization transfer,
the false asymmetries largely cancel in forming the helicity differences.

\begin{figure}[!t]
\begin{center}
\includegraphics[width=3.0in]{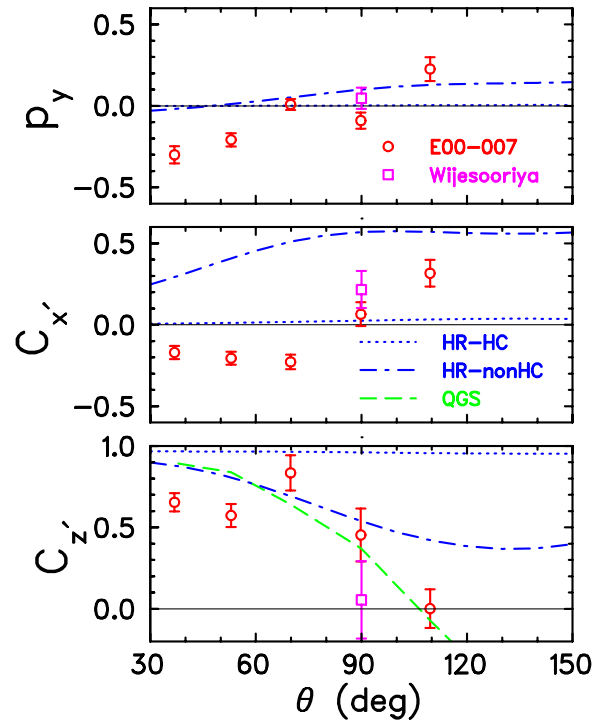}
\caption[]{\label{fig:data} (Color online) Polarization transfers $C^\prime_x$, $C^\prime_z$ 
and induced polarization $p_y$ in deuteron photodisintegration. 
Only statistical uncertainties are shown.
See text for details.}
\end{center}
\end{figure}

Figure~\ref{fig:data} compares
the proton recoil 
polarization of this work (E00-007), at $E_{\gamma}$ $\approx$ 2 GeV, 
with earlier results (Wijesooriya) \cite{wijesooriya01} 
at $E_{\gamma}$ = 1.86 GeV, and calculations.
A slow energy dependence of the recoil polarizations 
above $E_{\gamma}$ $\approx$ 1 GeV was found in
\cite{wijesooriya01}, and our new measurements at $\theta_{cm}$ = 90$^\circ$
are compatible with the earlier results.
All three polarization components are consistent with 
a smooth variation with angle, and with crossing zero
near $\theta_{cm}$ = 90$^\circ$.
Both $C_{x'}$ and $p_y$ start out negative and moderately sized
at forward angles, while $C_{z'}$ is positive and large.
As $p_y$ and $C_{x'}$ do not generally vanish, we again confirm
that HHC does not hold.

The longitudinal polarization is given by \cite{bara}
\begin{equation}
f(\theta) C_{z'} = \sum_{i=1}^6 \sum_{\pm}  \pm |F_{i,\pm}|^2,
\end{equation}
with $f(\theta)$ the cross section.
It is insensitive to phases of amplitudes.
Except for the negative signs, $C_{z'}$ would equal the cross
section, and could be predicted as reliably.
In contrast, $C_{x'}$ and $p_y$ are the real and imaginary 
parts of the same sum of interfering amplitudes \cite{gg02,bara}, and 
are highly sensitive to phases, and difficult to predict.

Two calculations of the spin observables are available.
Figure~\ref{fig:data} shows that the QGS model \cite{qgspol}
predicts a longitudinal polarization transfer in good qualitative agreement 
with the measured data, but makes no prediction for the transverse 
polarizations, due to their sensitivity to phases.
Given the good agreement with deuteron photodisintegration cross sections
in the few GeV region \cite{clas}, the QGS model must be 
regarded as the most successful existing model of photodisintegration at a few GeV.

Figure~\ref{fig:data} also shows predictions for all three observables
from the HR model \cite{hrmpol}.
It should be noted that these calculations are at the lower edge
of the nominal validity range of the model.
Also, since the $pn$ spin amplitudes are not well constrained by data,
the $pn$ amplitudes are based on $pp$ data.
Thus, there are large uncertainties in the predictions.
One calculation (dotted line) assumes that there is only small helicity
nonconservation, leading to small values of $C_{x'}$ and $p_y$,
and $C_{z'}$ being nearly unity~\footnote{
Helicity conservation leads to
$C_{x'}$ and $p_y$ vanishing, but the limit for $C_{z'}$ is model dependent.
In the HR model, the helicity conserving amplitude $F_{1+}$ $\geq$ $F_{5\pm}$,
while $F_{3-}$ is small, as a $+$ helicity photon 
scatters from a $+$ helicity proton. 
The limit is {\em not} the same as 
the usual one presented in \cite{gg02}, which assumes $F_{1+}$ $\approx$ $F_{3-}$.
}.
The second (dash-dot line) calculation assumes large helicity non-conservation.
The comparison with all observables supports large helicity nonconservation,
but clearly the predictions for the transverse polarization are
not in sufficiently good agreement.
However, ref.~\cite{hrmpol} points out that the transverse polarizations 
are approximately proportional to a particular amplitude in $pn$ scattering 
(``$\phi_5$'') that vanishes at $\theta_{cm}$ $=$ 90$^{\circ}$ in the isovector channel.
Thus, the transverse polarization data might be indicating that the isovector channel 
dominates over the isoscalar channel, more so than in the calculation.
This observation is consistent with the situation in the $\Sigma$ asymmetry, mentioned above.
While the HR model does not agree quite as well as the QGS model with $C_{z'}$
or with the cross sections for energies around 2 GeV, it is
at least as successful at predicting cross section above about 3 GeV
\cite{gg02}, and its $p_y$ predictions are consistent with the large-angle 
data.

To summarize, we provide new benchmark data for polarizations in
deuteron photodisintegration.
The two models which predict the longitudinal polarization transfer, 
the QGS and HR models, are in qualitative agreement with these data;
while neither model adequately explains the transverse polarizations, 
the HR model indicates the qualitative behavior might arise from isovector dominance.
High-energy photodisintegration of $pp$ pairs \cite{brodsky03} is 
the next major test of the underlying dynamics and of the 
theoretical models, as these two models give very different predictions
for $pp$ photodisintegration.

We thank Drs. Brodsky, Grishina, Hiller, Lee, Kondratyuk, Miller, Radyushkin, 
Sargsian and Strikman for many interesting discussions.
We thank the JLab physics and accelerator divisions for their support,
particularly the cryotarget and Hall A technical staffs for their extensive
support at short notice.
This work was supported by 
the U.S.\ National Science Foundation,
the U.S.\ Department of Energy,
and
the Italian Istituto Nazionale di Fisica Nucleare.
The Southeastern Universities Research Association (SURA) operates
the Thomas Jefferson National Accelerator Facility under DOE
contract DE-AC05-84ER40150.
The polarimeter was funded by the U.S.\ National Science
Foundation, grants PHY 9213864 and PHY 9213869.


\begin{thebibliography}{999}


\bibitem{alexa99} L.C.~Alexa {\it et al.}, \Journal{\PRL}{82}{1374}{1999}.
\bibitem{abbott00} D.~Abbott {\it et al.}, \Journal{\PRL}{84}{5053}{2000}.
\bibitem{benmokhtar05} F.~Benmokhtar {\it et al.}, \Journal{\PRL}{94}{082305}{2005}
\bibitem{rvachev05} M.~Rvachev {\it et al.}, \Journal{\PRL}{94}{192302}{2005}.

\bibitem{ne8} J.~Napolitano {\it et al.}, \Journal{\PRL}{61}{2530}{1988};
              S.J.~Freedman {\it et al.}, \Journal{\PRC}{48}{1864}{1993}.
\bibitem{ne17} J.E.~Belz {\it et al.}, \Journal{\PRL}{74}{646}{1995}.
\bibitem{89012} C.~Bochna {\it et al.}, \Journal{\PRL}{81}{4576}{1998}.
\bibitem{96003} E.C.~Schulte {\it et al.}, \Journal{\PRL}{87}{102302}{2001}.
\bibitem{99008} E.C.~Schulte {\it et al.}, \Journal{\PRC}{66}{042201R}{2002}.
\bibitem{clas} M.~Mirazita {\it et al.}, \Journal{\PRC}{70}{014005}{2004}.

\bibitem{ccr} S.J.~Brodsky and G.R.~Farrar, \Journal{\PRL}{31}{1153}{1973};
V.~Matveev {\it et al.}, Nuovo Cim. Lett.\ {\bf 7}, 719 (1973).
\bibitem{rossi05} P.~Rossi {\it et al.}, \Journal{\PRL}{94}{012301}{2005}.

\bibitem{gg02} R. Gilman and Franz Gross, \Journal{\JPG}{28}{R37}{2002}.

\bibitem{rna} S.J.~Brodsky and J.R.~Hiller, \Journal{\PRC}{28}{475}{1983}.
\bibitem{qgs} E.\ De Sanctis {\it et al.}, Few Body Syst.\ Suppl.\ {\bf 6}, 229 (1992); 
              L.\ A.\ Kondratyuk {\it et al.}, \Journal{\PRC}{48}{2491}{1993};
              V.Yu Grishina {\it et al.}, \Journal{\EPJA}{10}{355}{2001}.
\bibitem{hrm} L.L.~Frankfurt, G.A.~Miller, M.M.~Sargsian, and M.I.~Strikman, 
              \Journal{\PRL}{84}{3045}{2000}; 
              L.L.~Frankfurt, G.A.~Miller, M.M.~Sargsian, and M.I.~Strikman, 
              \Journal{\NPA}{663}{349}{2000}. 
\bibitem{ar}  A.\ Radyushkin, private communication.
\bibitem{ljd} B.~Julia-Diaz and T.-S.~H.~Lee, \Journal{\MPLA}{18}{200}{2003}.

\bibitem{qgspol} V.Yu Grishina {\it et al.}, \Journal{\EPJA}{18}{207}{2003};
                 V.Yu Grishina {\it et al.}, \Journal{\EPJA}{19}{117}{2004}.
\bibitem{hrmpol} M.M.~Sargsian, \Journal{\PLB}{587}{41}{2004}.

\bibitem{Adamian00}  F.~Adamian {\it et al.}, \Journal{\EPJA}{8}{423}{2000}.
\bibitem{wijesooriya01} K.~Wijesooriya {\it et al.}, \Journal{\PRL}{86}{2975}{2001}.

\bibitem{hhc}See S.J.~Brodsky and G.P.~Lepage, \Journal{\PRD}{24}{2848}{1981},
and references therein.

\bibitem{gpr} T.~Gousset, B.~Pire, and J.P.~Ralston, \Journal{\PRD}{53}{1202}{1996}.

\bibitem{hallanim} J. Alcorn {\it et al.}, \Journal{\NIMA}{522}{294}{2004}.

\bibitem{aalm} H.\ Olsen and L.\ C.\ Maximon, \Journal{\PR}{110}{589}{1958}.

\bibitem{punjabi05} V.~Punjabi {\it et al.}, \Journal{\PRC}{71}{055202}{2005};
                    Publisher's Note \Journal{\PRC}{71}{069902}{2005}. 

\bibitem{rcs05} D. J. Hamilton {\it et al.}, \Journal{\PRL}{94}{242001}{2005}.

\bibitem{wijesooriya02} K. Wijesooriya {\it et al.}, \Journal{\PRC}{66}{034614}{2002}.

\bibitem{jones00} M.K.~Jones {\it et al.}, \Journal{\PRL}{84}{1398}{2000};
\bibitem{gayou01} O.~Gayou {\it et al.}, \Journal{\PRC}{64}{038202}{2001}.
\bibitem{gayou02} O.~Gayou {\it et al.}, \Journal{\PRL}{88}{092301}{2002}.
\bibitem{hallahe403} S. Strauch {\it et al.}, \Journal{\PRL}{91}{052301}{2003}.
\bibitem{ndelta05} J. J. Kelly {\it et al.}, \Journal{\PRL}{95}{102001}{2005}.

\bibitem{akre}   A.I. Akhiezer and M.P. Rekalo, Sov. Phys. Doklady {\bf 13}, 572 (1968).
\bibitem{dombey} N. Dombey, Rev. Mod. Phys. {\bf 41}, 236 (1969).
\bibitem{akre2}  A.I. Akhiezer and M.P. Rekalo, Sov. J. Part. Nucl. {\bf 3}, 277 (1974).
\bibitem{acg}    R.~Arnold, C.~Carlson and F.~Gross, \Journal{\PRC}{23}{363}{1981}.

\bibitem{mcn} M.W.~McNaughton {\it et al.}, \Journal{\NIMA}{241}{435}{1985}.
\bibitem{pomme} N.E.~Cheung {\it et al.}, \Journal{\NIMA}{363}{561}{1995};
B.~Bonin {\it et al.}, \Journal{\NIMA}{288}{379}{1990}.

\bibitem{bara}V.P.~Barannik {\it et al.}, \Journal{\NPA}{451}{751}{1986}.

\bibitem{brodsky03} S.J.~Brodsky {\it et al.}, \Journal{\PLB}{578}{69}{2003};
 E.~Piasetzky {\it et al.}, Jefferson Lab proposal E03-101.


\end{thebibliography}
\end{document}